\documentclass {article}[11pt] 

\usepackage{fullpage}
\usepackage{color}
\usepackage{epsfig,subfigure}
\usepackage{epsfig}
\usepackage{latexsym}
\usepackage {amsmath}
\usepackage {stackrel}
\usepackage{psfig}

\newtheorem{definition}{Definition}

\newtheorem{lemma}{Lemma}
\newtheorem{corollary}{Corollary}

\newcommand{\todo}[1]{}

\newcommand{\cali}{{\cal I}}
\newcommand{\ith}{$i^{\mbox{\footnotesize th}}$}

\newcommand{\set}[1]{\left\{#1\right\}}

\newcommand{\iarrow}{{\stackrel {i} {\rightarrow}}}
\newcommand{\jarrow}[1]{{\stackrel {#1} {\rightarrow}}}
\newcommand{\ijarrow}{{\stackrel {i,j} {\rightarrow}}}
\newcommand{\level}{\ell}

\begin{document}
\title{A Generic Top-Down Dynamic-Programming Approach to Prefix-Free Coding\footnote{Work of all of the authors was partially supported by Hong Kong CERG grant 613507. Work of the 2nd 
and 3rd  authors also partially supported by a grant from the National Natural Science
Foundation of China (No. 60573025) and by the Shanghai Leading
Academic Discipline
Project, project number B114.
}}
\author
{Mordecai J. GOLIN\\
Dept of Computer Science\\
\quad  \& Engineering\\
Hong Kong UST\\
Hong Kong, China\\
{\em golin@cs.ust.hk}
\and 
Xiaoming XU\\
Dept of Computer Science\\
\quad  \& Engineering\\
IIPL, Shanghai Key Lab\\
 Fudan University\\
 Shanghai, China\\
 {\em xuxiaoming@fudan.edu.cn}
\and Jiajin YU\\
Dept of Computer Science\\
\quad  \& Engineering\\
IIPL, Shanghai Key Lab\\
 Fudan University\\
 Shanghai, China\\
 {\em jiajinyu@fudan.edu.cn}
}

\maketitle

\begin{abstract}
  Given a probability distribution over a set of $n$ words to be transmitted,
  the {\em Huffman Coding} problem is to find a minimal-cost prefix free code
  for transmitting those words.  The basic Huffman coding problem can be solved
  in $O(n \log n)$ time but variations are more difficult.  One of the standard
  techniques for solving these variations utilizes a top-down dynamic
  programming approach.

  In this paper we show that this approach is amenable to dynamic programming speedup
  techniques, permitting a speedup of an order of magnitude for many algorithms
  in the literature for such variations as mixed radix, reserved length and
  one-ended coding.  These speedups are immediate implications of a general structural 
  property that permits batching together the calculation of many DP entries.
  

\end{abstract}

\eject

\section{Introduction}
\label{sec:intro}

Optimal prefix-free coding, or \emph{Huffman coding}, is a standard compression
technique.  Consider an \emph{encoding alphabet} $\Sigma = \{\sigma_1, \ldots,
\sigma_r\}$.  A {\em code} $W=\{w_1,w_2,\ldots,w_n\}$ is a set of \emph{code
  words} $w_i \in \Sigma^*$. Code $W$ is \emph{prefix-free} if $\forall w,w' \in
W$ $w$ is not a prefix of $w'.$ 
As an example,  $\{01, 00, 100\}$ is a prefix-free code but
 $\{01, 00, 001\}$ is not, because $00$ is a prefix of $001$.
 
For $w \in \Sigma^*$, let $|w|$ devote the {\em
  length} of $w,$ i.e., the number of characters in $w.$  For example
  $|0101| =4.$

The input to the problem is a discrete probability distribution 
$P=\{p_1,p_2,\ldots,p_n\}$, $\sum_i p_i = 1,$ $\forall i,\, p_i \ge 0.$

The output is a a prefix-free code $W = \{w_1,w_2,\ldots,w_n\}$ whose expected
encoding length $\sum_{i=1}^n p_i |w_i|$ is minimized over all $n$ word
prefix-free codes.  Formally set $Cost(W,P) = \sum_{i=1}^n w_i p_i$.  Then
\begin{equation}
Cost(P) = \min_{\stackrel{W' \subseteq \Sigma^*,\, |W'| = n}{\mbox{\footnotesize $W$ is prefix-free}}} Cost(W',P)
\end{equation}

In \cite{Hu52}, Huffman gave a classical $O(n \log n)$ time greedy algorithm for
solving the binary case ($r=2$) of this problem.  Huffman also extended the
algorithm to solve the general $r$-ary case with the same time bound.  If the
$p_i$'s are given in sorted order, Huffman's algorithm can be improved to
$O(rn)$ time \cite{Va76}.

The correctness of the Huffman algorithm, although easy to prove, is very
strongly dependent upon properties of optimal prefix-free codes.  Almost any
extra constraint or generalization added to the problem description will
invalidate the algorithm's correctness.  Many such constraints/generalizations
appear in the literature (\cite{Abrahams01} is a nice survey) and all require
special purpose algorithms to address them.

Some examples of such prefix-free coding problems are
 \emph{Length-limited}  coding e.g, \cite{Ka61,LaHi90,AgSc94+,Sc98},
\emph{Unequal-cost} coding e.g., \cite{Blach-54,GR-98,BGLR-02,Dumit06}
\emph{Mixed-radix}  coding \cite{ChuGill92},
\emph{Reserved-length} coding \cite{baer-2007},
and   \emph{One-ended} coding \cite{BY-90,CSP-94,ChanG00},

The major observation is that all of the best algorithms known for these
problems use some form of dynamic programming (DP) to build an optimal
(min-cost) coding tree that corresponds to an optimal code.

These DPs primarily differ in whether they build the tree from the bottom-up or the
top-down. 
The best algorithms for 
Length-limited and Unequal-cost coding
 use what essentially reduces to a bottom-up DP model 
combined with some DP-speedup
techniques, e.g., Monge speedups using the SMAWK algorithm of \cite{AgKl87+}
(see \cite{BGLR-02} for an example of this technique and \cite{Sc98} for
a  more sophisticated but more specialized speedup method);
the best algorithms for  Reserved-length and One-ended coding
use a top-down DP approach. (Mixed-radix coding \cite{ChuGill92} uses a totally different
DP approach described later)

Length-Limited and Unequal-Cost coding
{\em could} be 
solved using a top-down approach but the bottom-up solutions are better for two
reasons.  The first is that the bottom-up solutions use a more compact solution
space than corresponding top-down ones would. This is due to the exploitation of
some very problem-specific combinatorial structures of their
corresponding optimal code trees.  The
second is that their bottom-up DPs turn out to have special properties,
e.g., the Monge property, 
which enable speeding up the calculation of table entries.
The top-down DPs used in the last two problems don't have such a compact representations
and they also,  before this paper, didn't seem to possess  any special property
that would lead to speedups.

The main result of this paper is a revisiting of the  
generic  top-down DP approach
for solving prefix-free coding problems.  We will show that, in this setup, many natural coding
problems will  have an obvious and simple batching speedup.  That is,  we will
be able to partition the DP table entries  into smaller batches (groups)  and 
exploit relationships between entries  within a batch  to 
fill in all of the entries in each batch in $O(1)$ amortized time per entry.
This will enable
 speeding up  known solutions to  the last three problems
by at least one factor of $n$. The interesting observation is
that the {\em same} speedup technique works for all of these problems.
Table \ref{tab:Results} lists the speedups.


\begin{table}[t]
\label{tab:Results}
\begin{center}
\begin{tabular}{||l|c|c||} \hline
Problem & Previous Best  Result  & This paper \\[.02in] \hline \hline
Mixed Radix Coding & $O(n^4 \log n)$ \cite{ChuGill92} & $O(n^3)$\\[.02in] \hline
Reserved Length Coding:  & &  \\
$g$ specific lengths given & $O(gn^3)$ \cite{baer-2007} &$O(gn^2)$ \\[.02in] \hline
Reserved Length Coding: & &  \\
at most $g$ lengths allowed & $O(g^3 n^3 \log^g n)$ \cite{baer-2007}& $O(g n^2 \log n)$ \\[.02in] \hline
One-ended Coding & $O(n^3)$ \cite{ChanG00}& $O(n^2)$\\[.02in] \hline
\end{tabular}
\end{center}
\vspace*{-.1in}
\caption{A comparison of our new results with previous ones}
\end{table}

\subsection{The problems}
 
\begin{figure}[t]
\centerline{\epsfig{file=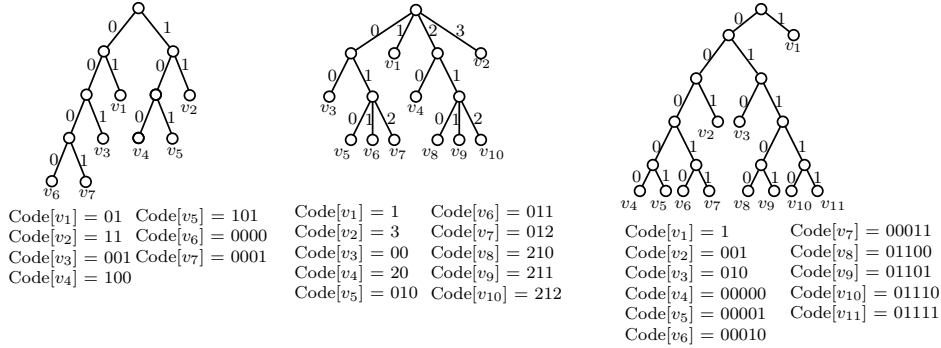, width= 5.0 in}}
\vspace*{-.1in}
\caption{Examples of the code-tree correspondence.  Codes are written below their
corresponding tree.  The leftmost figure is a standard binary tree.  The middle is
a Mixed-radix tree with level arities $(t_0,t_1,t_2) = (4,2,3).$  The rightmost is
a Reserved-length tree with codewords only on levels $\Lambda=\{1,3,5\}$. 
Note that leaves in all the trees are labelled from top-to bottom}
\label{Fig:Tree_rep}
\end{figure}

 
 We start by quickly recalling the standard correspondence
between prefix-free codes and trees.  Let $r = |\Sigma|$ be the size of the
alphabet and consider an $r$-ary tree $T$ in which the \ith edge leaving a node
is labelled with character $\sigma_i$.  Associate with node $v \in T$ the unique
word read off walking down the path from the root to $v.$ The set of words $W$
associated with the leaves of $T$ is prefix-free.  Conversly, given a
prefix-free code $W$ one can build a tree whose leaves are exactly the nodes
associated with the words of $W$.  See Figure \ref{Fig:Tree_rep}

Given tree $T$ associated with code $W$, denote its leaves by
$v_1,v_2,\ldots,v_n$ where $v_i$ is the leaf associated with word $w_i.$ Let
$d(v_i)$ be the {\em depth} of $v_i$ in $T.$ By the correspondence, $d(v_i) =
|w_i|$ so
$$Cost(T,P) = \sum_i d(v_i) p_i = \sum_i |w_i| p_i = Cost(W,P)$$
where $Cost(T,P)$ can be understood as the weighted external-path-length of $T.$
So the prefix-free coding problem is equivalent to finding a tree with minimal
external path length.  For this reason, most algorithms for finding prefix-free
codes are stated as tree algorithms.

We now quickly discuss the problems mentioned in the previous sections and
their tree equivalents and then  state our new results for these problems.

\medskip

\par\noindent\underline{\em Mixed-Radix Coding:}\\
In Mixed-Radix Coding the  size of the encoding alphabet used depends upon
the position of the character within the codeword.  This corresponds to
constructing a tree in which the {\em arity} (number of children) of an internal
node depends upon the level of the node. That is, as part of the problem
definition, we are given a sequence $t_0,t_1,\ldots$ of integers, $t_i \ge 2$
such that the maximum arity of a node on level $i$ is $t_i.$

The coding version of the problem was motivated \cite{ChuGill92} by coding with side-channel
information and the tree version by problems in multi-level data storage.  
Chu and Gill \cite{ChuGill92} solved this problem by introducing an
alphabetic version of it and then solving a special case of the alphabetic
version.  
Their algorithm runs in $O(n^4 \log n)$ time; we will  improve this to $O(n^3)$.

\smallskip
\par\noindent\underline{\em Reserved-Length Coding:}\\
Recall that $|w_i|$ is the length of the \ith codeword.  In reserved-length
coding there are specific restrictions as to the  permitted values of $|w_i|$.  There
are two versions of this problem.  In the first version, the {\em given-lengths case},
$\Lambda =
\{\gamma_1,\gamma_2,\ldots, \gamma_g\}$ is given as part of the input and we
must find a minimum-cost code such that $\forall i, |w_i|\in \Lambda$.  This
corresponds to building a min-cost tree in which all leaves are on levels in
$\Lambda.$ 

In the second version, the {\em $g$-lengths case},
 $\Lambda$ is not given in advance.  The restriction now is to find a minimum-cost code
under the restriction that $|\Lambda| \le g$, where $\Lambda$ is the set of codeword
lengths used.  This corresponds to building a tree in which at most $g$ levels
may contain leaves.


Baer \cite{baer-2007} introduces these problems in the context of fast decoding and
used a top-down DP approach to solve the first one in $O(|\Lambda| n^3)$ time 
and the second one in $O(n^3 g^3 \log^g n)$ time .  We will 
 reduce these two cases, respectively,
 to $O(|\Lambda| n^2)$ and $O(n^2 g \log n)$ time.

\smallskip
\par\noindent\underline{\em One-Ended Coding:}\\
In One-Ended coding the aim is to find a min-cost binary prefix-free code in
which every word must end with a {\bf 1}.  This corresponds to finding a
min-cost tree in which only right leaves (leaves that are the right children of
their parents) are labelled with the $p_i$ and counted in the calculation of the
cost.  

One-Ended Coding was introduced by  Berger and Yeung
\cite{BY-90} in the context of
self-synchronizing codes.  Their algorithm  ran in exponential time.
This was later improved  by De Santis, Capocelli and Persiano  \cite{CSP-94} to another exponential-time algorithm
with a smaller exponential base.  Chan and Golin \cite{ChanG00} showed how to use top-down DP
to derive an $O(n^3)$ time algorithm.  We will  reduce
this down to an $O(n^2)$ time one. 

\smallskip


In Section \ref{Sec:topdown} we introduce a new coding problem 
called  {\em Generalized Mixed-Radix Coding}, develop a top-down DP approach for solving it
and then speed it up by  batching.  In Section \ref{sec:MRCandRLC} we 
reduce both Mixed-Radix Coding and Reduced Length Coding to (multiple) applications of
GMR and thus take adavantage of the DP speedup. In Section \ref{sec:One_Ended} we
reduce the running time of One-Ended coding using an almost identical technique.
Since the analysis of One-Ended coding is very similar to that of the GMR problem,
we do not provide the details in this extended abstract (but they are available in the appendix).

\section{The Top Down DP for Generalized Mixed-Radix Coding}
\label{Sec:topdown}

We start by  introducing the Generalized Mixed-Radix (GMR) problem, 
develop a top-down DP for solving it and 
then see how to speed it  up.  

\begin{figure}
\centerline{\epsfig{file=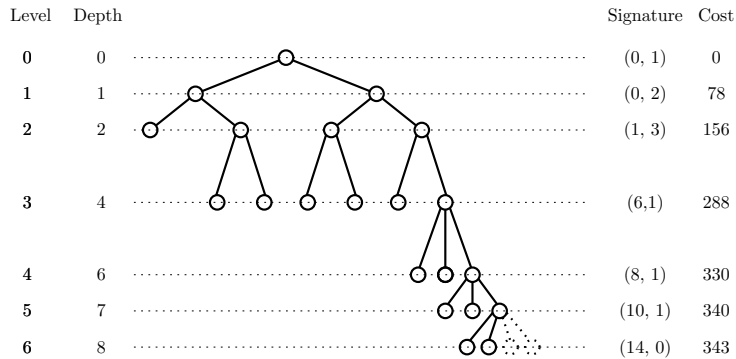, width= 3.8 in}}
\caption{A generalized mixed radix tree with $R=(2,2,2,3,3,4)$ and
$C=(1,1,2,2,1,1)$.   Note that $n=12$
but to make the tree full we needed to add two extra leavs on the bottom level.
Also, the signatures and costs on level $i$ are of the truncated tree containing the
nodes on the first $i$ levels. Costs are for $P=\{1,2,3,4,5,6,7,8,9,10,11,12\}$.}
\label{fig:GMR}
\end{figure}

In a \emph{generalized mixed radix} tree, both the \emph{arity} of an internal
node $v$ and the \emph{length} of an edge leaving $v$ depends on the {\em level} of
$v$. Figure \ref{fig:GMR} illustrates these and other definitions in this section.
More formally,

\begin{definition}
 Given a a sequence of arities $R=(r_1,r_2, \ldots)$ and a sequence
of edge length $C=(c_1, c_2 \ldots)$, a generalized mixed radix  (GMR) tree $T$
\emph{satisfying} $R$, $C$ is a tree in which internal node $v$ at
level $i-1$ has at most $r_i$ children and 
and the length of an edge from $v$ to any of its children is $c_i$.
\end{definition}

We now distinguish between the level $\ell(v)$  of a node v, which is the number of edges
from the root to $v$, and its depth,  which is the weighted path
length from the root to $v$. 
 More formally,
\begin{definition}
The level of node $v$ in tree $T$ is the number of edges on the unique path from the root to $v$ and will be denoted by $\level(v).$ The level  of tree $T$ is
$$\level(T) 	= 	\max\{ \level(v) \,\mid\, \mbox{$v$ is a node in $T$}\}
						=		\max\{ \level(v) \,\mid\, \mbox{$v$ is a leaf in $T$}\}.
$$

The {\em depth} of  a node  on level $i$ of the tree will be the sum of the lengths of
the edges on the path from the root to level $i$, i.e., $depth(v) = L(i) = \sum_{j=1}^{i} c_i.$
The depth of tree $T$ will be $depth(T) = L(\level(T)).$ 
\end{definition}

There is an obvious definition of cost in such trees.

\begin{definition}
Given $R,C$ as above	

Let $T$ be any generalized mixed radix tree $T$ for $R,C$ 
with $n$ leaves labeled $v_1, v_2, \cdots, v_n$.  Then
\begin{equation}
\label{eq:L_cost_def}
Cost(T)=\sum_i depth(v_i) \cdot p_i
\end{equation}
\end{definition}

The problem  to be solved is, given $P$ and $R,C$, find a min-cost tree $T^*$ with $n$
leaves,  i.e., 
$Cost(T^*)=min \left\{Cost(T) \mid \textit{T has $n$ leaves}
  \right \}$

\subsection{The Basic Top-Down Dynamic Program}
\label{subsec:Basic TD DP}

In this section we 
quickly describe  the standard top-down DP formulation.
Since variations of this formulation have been extensively used before for
various coding problems, 
e.g. \cite{GR-98,DolevKY99,ChanG00,baer-2007}, we only sketch
the method  but do not rigorously prove its correctness.


In what follows $P=\{p_1,p_2,\cdots ,p_n\}$ with $p_1 \ge p_2 \ge \cdots \ge p_n\ge 0$
is given and fixed.  The $p_i$ can be arbitrary weights and are not required  to sum to $1$.
The $p_i$ sequence is implicitly padded so that, for $i > n,$ $ p_i=0.$ 
Finally,  for $m \ge 0$, set $W_m = \sum_{i > m} p_i.$

We start with some standard simplifying assumptions  about min-cost (optimal)    trees $T^*$.
We will
show that there always exists at least one optimal tree satisfying these assumptions.  Since
our goal is to find any optimal tree, our search can be restricted to trees satisfying the
assumptions.  

In what follows, an internal  node $v$,  $\level(v) = i$, in tree $T$ will be {\em full} if
and only if it has $r_{i+1}$ children in $T.$  Tree $T$ will be {\em full} if
all of its internal nodes are full.


\medskip

\par\noindent{\em Assumption 1:}  If $i < j$ then, in $T^*$,  $depth(v_i) \le depth(v_j)$.\\
If this was not true   we could label $v_i$ with $p_j$ and $v_j$ with $p_i$.  The resulting
tree has cost no greater than the original one so it remains optimal.
We may therefore always implictly assume that leaf weights in trees are non-increasing
in the level of the tree.

\medskip
\par\noindent{\em Assumption 2:}  There  is a full tree $T^*$  with the same cost
as the optimal tree for $n$ leaves.  $T^*$ has $n'$ leaves where
$$\max(n,r_{\ell(T^*)}) \le n' \le n +r_{\ell(T^*)}-1$$

This will be a consequence of the padding of $P.$

Let $T$ be an optimal tree with {\em exactly} $n$ leaves.
Suppose  $\level(T)= \level(v_n) = \ell$.

First note that all internal nodes $v$ with $\ell(v) < \ell-1$ are full.  Otherwise
we could add a new leaf child of $v$ at level $\ell(v) + 1 < \ell(v_n)$ and
label it with $p_n$, creating a tree with smaller cost than optimal tree $T.$
Thus,  the only non-full internal nodes in $T$ are on level $\ell-1.$

Next note that we may assume that {\em at most} one 
internal node at level $(\ell-1)$ is not full;  otherwise leaves on level $i$ 
can be shifted to the left, 
so that all internal nodes on level-$(i-1)$, except possibly
the rightmost one,  are full. Make this rightmost node full by adding an appropriate number
 leaves to it and call this new full tree $T^*$.
Note that  $\ell(T^*) = \ell(T) = \ell$ and $cost(T) = cost(T^*).$ This 
follows because  $P$ was padded by setting 
$p_i=0$ for $i >n.$ Let $n'$ be the number of leaves in $T^*$.

By definition $n \le n'.$  Since $T$ contains at least one internal node on level $\ell-1$,
$T^*$ contains at least $r_{\ell}$ leaves on level $\ell-1.$  Thus
$\max(n,r_{\ell}) \le n'$

Furthermore, $ n' \le n + r_{\ell}-1$ 
because  $T^*$ was created by adding at most $r_{\ell}-1$ leaves to $T$.

\medskip
\par\noindent{\em Assumption 3:} $\level(T^*) \le n$.\\
This will follow from the fact that we may assume that $\forall i, r_i \ge 2.$

\medskip
\par\noindent{\em Assumption 4:}  The cost of  a tree is fully determined by its {\em leaf sequence},  i.e., the number of
leaves on each level.  No other structural properties need to be maintained.

This follows directly from the previous assumptions, i.e., the fullness of optimal trees. 
Although we will talk about constructing {\em trees} we
will really be constructing the corresponding
{\em leaf sequences,} e.g., sequences denoting how many leaves are on on each level. 
Since the cost of a tree is fully determined by its leaf sequence this does not 
cause any problems.

\bigskip
%


Suppose that $T'$ is a tree with $n' >n$ leaves. Create $T''$ by pruning the deepest
leaves from $T'$ one-by-one, until exactly $n$ leaves remain.  Then,
by construction, $T''$ is a tree with exactly $n$ leaves such that
$cost(T'') \le cost(T').$  This observation,  Assumption 2
and Assumption 3,
tell us that we can find the optimal tree for $n$ leaves by first finding
--  for
every $\ell \le n$,  and
every $n'$ satisfying $\max(n,r_{\ell}) \le n' \le n +r_{\ell}-1$--
the cost of the min-cost full tree of level at most $ \ell$ with
$n'$ leaves.  Then we take the minimum cost tree among all such trees and prune it until
it has exactly $n$ leaves.  The resulting tree wil be the optimal tree for $n$ leaves.

There are only  $O(n^2)$ $(\ell,n')$ pairs that need to be examined; given their costs,
finding the optimal pair requires only $O(n^2)$ time.  (The subsequent
pruning operation can easily  be done in $O(n)$ time.)
The hard part is, for each given $(\ell, n')$ pair, to find the costs of the appropriate
min-cost full tree and then, if necessary, build it.

The intuition behind the solution is to build  optimal  trees top-down,  
starting with an initial tree -- the root -- building successively
bigger   trees level-by-level by making some nodes on the bottom level internal.
Part of the specification of these intermediate
 {\em truncated trees} will be an explicit statement of the number of nodes on their bottom
 levels that will become internal when they are further expanded.  
The process  ends  when a tree whose bottom level contains no internal nodes
is constructed.

Since we are only interested in constructing full trees and the 
truncation of a full tree up to any level is full, 
{\em this process may implicitly assume that every intermediate tree built is full}.

To transform this intuition into a dynamic program  we will need 
to somehow encode the space of intermediate  trees compactly and introduce an appropriate definition
of {\em cost} for  intermediate trees.

\begin{definition}\ \\
\label{def:signature}
  Tree $T$ is an \emph{$i$-level tree} if all nodes $v\in T$ satisfy
  $\level(v) \le i$.
  
  If $T$ is  an $i$-level tree its  \emph{$i$-level signature}  is the  ordered
  pair 
  $$sig_i(T)=(m,b)$$ 
  in which 
  \begin{eqnarray*}
  m &=& |\left\{v\in T\,\mid\, \textit{v is a leaf, } \level(v) \le i\right\}|\\
  b &=& |\left\{v\in T\,\mid\, \textit{v is an internal node } \level(v) = i\right\}|
  \end{eqnarray*}
In the above definition, ``internal'' means  that the leaf at the bottom level 
is tagged as being made internal if the tree grows to its next level.

  Let $T$ be an $i$-level tree with $sig_i(T)=(m,b)$.  The $i$-level
\emph{partial cost} of $T$ is
\begin{equation}
\label{eq:partial-cost}
Cost_i(T)=\sum_{t=1}^m depth(v_t)\cdot p_t+ L(i)\cdot\sum_{t=m+1}^n p_t
\end{equation}
  
\end{definition}
Figure \ref{fig:GMR} illustrates this definition.

Consider the possible signatures that could occur.
Suppose that $T$ is an $i$-level  truncation of some optimal tree $T^*$ with $\level(T^*)=\ell$ and 
$sig_i(T)=(m,b)$.  

If $T$ is a proper truncation of $T^*$, i.e,  $b >0$,
then $\level(T) < \ell $.  Thus
every labelled leaf in $T$ is one of $v_1,\ldots,v_n$  in $T^*$ and 
every one of the $b$ nodes on the bottom level of $T$ that is labelled as ``internal'' is the
ancestor 
of one of $v_1,\ldots,v_n$ in $T^*.$  Thus $m+b \le n.$

If $T$ is not a proper truncation of  $T^*$, i.e., $T=T^*$,  then
$b=0$ and,  from Assumption 2,  
$\max(n,r_{\ell}) \le m \le n +r_{\ell}-1.$
This motivates 
\begin{definition}
\label{def:valid_1}
$(m,b)$ is a {\em valid $i$-level signature} if
\begin{itemize}
\item If $b >0$ then $m+b \le n.$
\item If $b=0$ then $\max(n,r_{i}) \le m \le n +r_{i}-1.$
\end{itemize}
\end{definition}
Obviously,  the number of valid $i$-level signatures is $O(n^2)$.

We can now introduce the  DP table.
\begin{definition}
\label{def:OPT}
  Let $(m,b)$ be a valid $i$-level signature. Set
  $OPT^i[m,b]$ to be the minimum $i$-level partial cost over all $i$-level trees $T$
  with signature $(m,b)$. More precisely
  \begin{equation}
  OPT^i[m,b] =  min\left\{Cost_i(T)\,\mid\, \exists T, \textit{T is a i-level full tree
      with}\ sig_i(T)=(m,b)\right\}
  \end{equation}
  If no such tree exists,  set $OPT^i[m,b] = \infty.$
\end{definition}  
If $t > n$ then $p_t=0$.  Thus, if $T^*$ is an  $i$-level full tree with $n' \ge n$ leaves then, by
definition, 
$Cost_i(T^*) = \sum_{t=1}^n depth(v_t) \cdot p_t = Cost(T^*).$
So, $OPT^i[n',0]$ will be the optimal actual cost of an
$i$-level full tree with $n' \ge n$ leaves, which is what we want.


\begin{definition}
\label{def:expand} 
Let $T'$ be an $(i-1)$-level tree with
 $sig_{i-1}(T')=(m',b')$. \\
Expand $T'$ to a full  $i$-level tree by adding all $r_{i} b'$ nodes
on level $i$.

For $b$ 
satisfying $ 0 \le  b \le b' r_{i}$, the 
{\em $b^{\mbox{\footnotesize{th}}}$ expansion} of $T'$ is the
$i$-level tree created by denoting $b$ of these $r_{i} b'$ nodes
as internal and
making  the remaining  $ b' r_{i} - b$ nodes into 
leaves. We denote this by
$$T = \textit{Expand}_i(T',b).$$
Note that $sig_i(T) = (m,b)$ where $m = m' + b' r_{i} - b$.
\end{definition}

We now extend the definition of expansions to signatures
\begin{definition}
\label{def:sig_exp}
If $(m',b'),$  $(m,b)$ are valid signatures such that  $ 0 \le  b \le b' r_{i}$
and $m = m' + b' r_{i} - b$ 
we write
$$(m',b') \iarrow (m,b).$$
\end{definition}

It is easy to prove the following by construction:
\begin{lemma}
\label{lem:sig_conv}
Let $T'$ be an $(i-1)$-level tree with $sig_i(T') = (m',b')$ and 
$(m,b)$ such that $(m',b') \iarrow (m,b).$ Then $T=Expand_i(T',b)$ is 
a well-defined $i$-level tree with
$sig_i(T) = (m,b)$.
\end{lemma}
This implies the following corollary, which is  the basis of the correctness of the dynamic program.

\begin{corollary}
\label{cor:path}
Let $(m_0,b_0) = (0,1)$ be the unique level-$0$ tree with internal root.  
Then the lemma implies that every sequence 
\begin{equation}
\label{eq:sequence}
(m_0,b_0) \jarrow 1
  (m_1,b_2) \jarrow 2
  \cdots	\jarrow {i-1}
  (m_{i-1},b_{i-1}) \jarrow {i}
  (m_i,b_i)
\end{equation}
 corresponds to an  $i$-level tree $T$ with $sig_i(T) = (m_i,b_i)$
 that can be constructed top-down from the root 
 by following the appropriate expansions given by the  sequence.
 In particular,  if $(m_i,b_i) = (n',0)$ then the constructed tree has
 exactly $n'$ leaves.
\end{corollary}

\begin{lemma}
\label{lem:Sum}
Let $T'$ and $T$ be, respectively, $i-1$ and $i$-level trees with
$sig_{i-1}(T') = (m',b')$ and $sig_i(T) = (m,b)$.  If  $(m',b') \iarrow (m,b)$ then
$$Cost_{i}(T) = Cost_{i-1}(T') + c_iW_{m'}.$$
\end{lemma}
\par\noindent{\em Proof: } 
  From Lemma \ref{lem:sig_conv},  $T= Expand(T',b)$ so level $i$ contains
  the leaves $v_{m'+1},\ldots,v_m$  and 
\begin{equation}\nonumber
  \begin{split}
    Cost_i(T) & = \sum_{t=1}^m depth(v_t)\cdot p_t+ L(i)\cdot\sum_{m<t\le
      n} p_t\\
    & = \sum_{t=1}^{m'} depth(v_t)\cdot p_t+ (L(i-1) + c_i) \left( \sum_{m'<t \le
      m} p_t +\sum_{m<t\le n} p_t\right)\\
    & =Cost_{i-1}(T')+ c_i\sum_{m'<t\le n} p_t = Cost_{i-1}(T')+ c_i W_{m'}
  \end{split}
\end{equation}
This tells us that the cost of the $n'$-leaf tree associated with sequence
(\ref{eq:sequence}) can be calculated level by level to be
$\sum_{t=1}^{i} c_t W_{m_t}$ where ${m_i} = n'.$ 
Combining all of the above,  we can now write a simple DP that models building
optimal trees from the top-down.
\begin{lemma}
\label{lem:DP_basic}
The optimal cost of an $i$-level tree with signature $(m,b)$ 
satisfies
\begin{equation}
\label{eq:basicDP}
OPT^i[m,b] = 
\min_{ {\tiny \{(m',b') \,\mid\,  (m',b') \iarrow (m,b)\}} }
		\Bigl\{ OPT^{i-1}[m',b']+ c_i W_{m'}\bigr\}.
\end{equation}
Initial conditions are that $OPT^0[0,1] =  0$ with all
other entries being set to $\infty$.		
\end{lemma}

The entries $OPT^i[m,b]$ only depend upon the entries $OPT^{i-1}[m,b]$ so
the table can be filled in using  the order  $i=1,2,\ldots,n.$ 

For any given level $i$ there are only $O(n^2)$ valid $i$-level signatures.
From Definition \ref{def:sig_exp},  for every $(m,b)$, there are only
 $O(n)$  signatures such that
$(m',b') \iarrow  (m,b)$  So, for fixed $i,$  
filling in all of the  $OPT^i[m,b]$
requires $O(n^3)$ time,  with $O ( n^4)$ total time needed to fill in all of the enties
 $OPT^i[m,b]$, $i \le n$.

We will now see how, for fixed $i$,  to calculate the values $OPT^{i}[m,b]$ in
only $O(n^2)$ time.  This will, as promised,  reduce the total running time for filling
in all $n$ levels of the table to to $O(n^3)$.

\vspace*{-.1in}
\subsection{Batching for Speedup}
\label{subsec:batching}
We now see how to fill in the DP entries in a faster way.  We first need two more definitions.
\begin{definition}
\label{def:old and new sigs}
For $1< d $, define
$$
\cali (d)  =  \left\{ (m,b)   \,\mid\, m+ b=d\right\},
\qquad
\cali'_i(d) = \left\{ (m',b') \,\mid\, m'+b'r_{i}=d\right\}.
$$
\end{definition}
For any fixed $i$ and  $(m,b) \in \cali_i(d)$, 
the definition of $\iarrow$ implies that \\[.01in]
\centerline{
``$(m',b') \iarrow (m,b)$'' \quad if and only if \quad
``$(m',b') \in \cali'_i(d)$ with $b \le b' r_i$''.
}\\[.02in]
This immediately permits rewriting (\ref{eq:basicDP})
as
\begin{lemma}
\label{lem:DP_batched}
If $(m,b) \in \cali(d)$ for some $d \le n$ and
\begin{equation}
\label{eq:batched1}
OPT^i[m,b]  =
\min_{ \stackrel {(m',b') \in \cali'_i(d)} {b \le b' r_{i}} }
		\Bigl\{ OPT^{i-1}[m',b']+ c_i W_{m'}\Bigr\}.
\end{equation}
\end{lemma}

We now claim that, for fixed $d \le n$, the calculation of the values 
 $OPT^i[m,b]$ for {\em all} $(m,b) \in \cali(d)$ can be batched together in 
 in $O(d)=O(n)$ time, i.e., in amortized $O(1)$ time per entry.

For fixed $d\le n$ suppose $(m',b') \in \cali'_i(d)$, i.e., $m' + b' r_{i} = d.$
This implies $b' \le \lfloor d/r_{i}\rfloor$.  Set 
$$\forall 0 \le b' \le \lfloor d/r_{i}\rfloor,\quad 
\gamma(b') 
= OPT^{i-1}[m',\,b'] + c_i W_{m'}
= OPT^{i-1}[d- b' r_{i},\,b'] + c_i W_{d- r_{i}b'}.
$$
These can be precalculated in $O(d)$ time.
Then (\ref{eq:batched1}) just says that for $(m,b) \in \cali(d)$, 
\begin{equation}
\label{eq:gamma_def}
OPT^i[m,b] = \min \left\{ \gamma(b') \ \Big|\   (b/ {r_{i}}) \le b' \le \left\lfloor  d/{r_{i}}\right\rfloor \right\}.
\end{equation}

$(m,b) \in \cali(d)$ implies $m = d -b$.  Since $b \le r_{i} \lfloor d/r_{i}\rfloor$, 
$ m \ge t = d - r_{i} \left \lfloor  d/{r_{i}} \right \rfloor = d \bmod r_{i}.$
So
$$\cali(d) = \left\{\, X_m \,|\,  m = t, t+1, \ldots ,d-1\right\}
\quad\mbox{where}\quad
X_m = \bigl(m,\,  d-m\bigr).
$$
%
Then 
(\ref{eq:gamma_def}) can be rewritten as 
\begin{equation}
 \label{eq:X_OPT}
OPT(X_m) = \min \left \{ \gamma(b') \ \big|\  ({d-m})/{r_{i}}  \le b' \le \left \lfloor  d/ {r_{i}}\right \rfloor \right\}
\end{equation}
This immediately yields 
\begin{eqnarray*}
OPT[X_t] &=&   \gamma\left(\left \lfloor  d /{r_{i}}\right \rfloor\right) \\[0.05in]
\forall m >t,\ OPT[X_{m}] &=& 
	\left\{
		\begin{array}{ll}
			OPT[X_{m-1}] & \mbox{if $r_{i} \not | (d-m)$}\\[0.08in]
			\min\left( OPT[X_{m-1}], \gamma\left(  ({d-m})/{r_{i}}\right)\right)
				& \mbox{if $r_{i}  | (d-m)$}
		\end{array}
	\right.
\end{eqnarray*}
Thus, we can calculate all of the $OPT[X_m]$ for $X_m \in \cali(d)$ 
in $O(d)$ time by working in the order
$m=t,t+1,\ldots,d.$

For fixed $i$, to fill in all valid signatures  we start by implicitly setting  all  
entries to $\infty$ and then iterate for $d=1,2,3,\ldots$, for each value of $d$ using $O(d)$ time
to calculate all of the entries $OPT^{i}[m,b]$ with $m+b =d$.  The question is where to
stop the iteration.  

From Definition \ref{def:valid_1}  we know that if $b > 0$ then $d = m+b \le n$ while if $b=0$ then
$\max(n, r_i) \le m \le n + r_i -1.$  There are now two cases.  

\begin{itemize}
\item [\bf $r_i \le n:$] 
Then $m+b < 2n$ so stop the iteration at $d = 2n-1$. All of the valid entries will
have been filled in using $\sum_{d < 2n} O(d) = O(n^2)$ time.
\item [$r_i > n:$]  In this case if $b=0$ then $r_i \le m \le r_i + n-1$ and 
$(m',b') \iarrow (m,0)$ implies  that either $(m',b') = (m,b)$ or
$(m',b') = (m-r_i,1),$ i.e., $(m,0)$ has only two possible predecessors.
 We can therefore fill in the full table in two phases.  In the first,
fill in all valid entries $OPT^{i}[m,b]$ with $m+b =d$ for $d=2,3,\ldots, n$ in $O(n^2)$ time.
In the second,  fill in the all valid entries of the form $OPT^i[m,0]$ with $r_i \le m \le r_i + n-1$
in $O(n)$ time, by checking the two predecessors of each possible entry.
\end{itemize}

\vspace*{-.2in}
\section{Mixed-Radix Coding and Reserved-Length Coding}
\label{sec:MRCandRLC}
We now see  how to solve both Mixed-Radix Coding and Reserved-Length Coding via the GMR approach.

\vspace*{-.1in}
\subsection{Mixed-Radix Coding}
Chu and Gill's  Mixed-Radix Coding problem \cite{ChuGill92}  is exactly the
GMR problem restricted to  all of the edge costs being identically 1, i.e., $\forall i, c_i=1.$ 
The algorithm in the previous section solves this in $O(n^3)$ time, improving upon the
$O(n^4 \log n)$ time of \cite{ChuGill92}.

\subsection{Reserved-Length Coding}

\begin{figure}[t]
\centerline{\epsfig{file=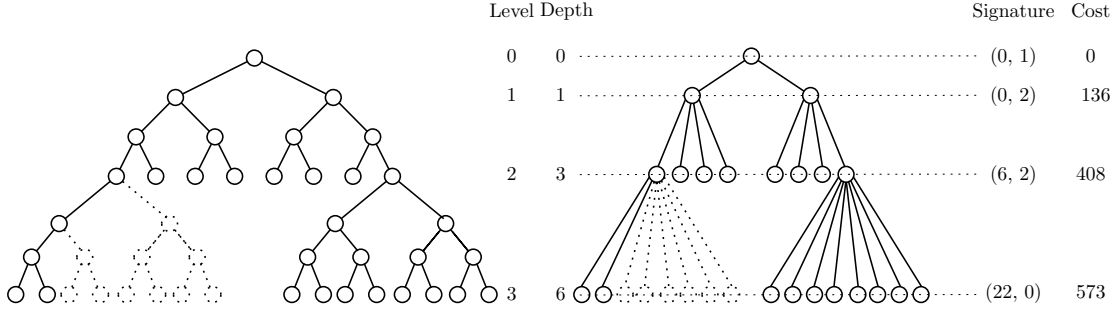, height= 1.6 in}}
\caption{A Reserved-Length tree (left) with $\Lambda=\{1,3,6\}$, $n=16$ leaves
and the corresponding GMR tree (right).  Note that even though they are allowed,  there are no leaves on level 1. All internal nodes on level $3$ should have 8 descendents on
level $6$ so  6 leaves were added to make the tree full.  Signatures and costs for the GMR tree
at level $i$ are for the truncated tree containing the first $i$ levels. Costs are for $P=\left\{1,2,\ldots, 16\right\}$.}
\label{Fig:Reserved} 
\end{figure}

In the {\em reserved-length coding problem}, there are restrictions as to permissible 
codeword lengths.  In the tree version of the problem,  these become restrictions on
the allowable levels on which leaves can appear.  More formally. let 
$$Level(T) = \{\level(v) \,|\, \mbox{$v$ a leaf in $T$}\}.$$
There are two versions of the problem.  

In the first version of the problem, the 
{\em given-lengths} case,  a  set of integers 
$\Lambda = \set{ \gamma_1, \ldots, \gamma_g}$ is given (w.l.o.g, $0 < \gamma_1 < \gamma_1 <
\gamma_2 < \cdots < \gamma_g$; we may  also add $\gamma_0=0$ since, if $n >1,$
the root will never be internal) and  we are 
 asked to find a minimum-cost $r$-ary tree among
all trees with $Level(T) \subseteq \Lambda$.

In the second version of the problem,  the
{\em $g$-lengths} case,  an integer $g$ is given and we are asked to find
a minimum-cost $r$-ary tree among
all trees with $|Level(T)| \le g.$

\subsubsection{The Given-Lengths Case}
\label{sec:given-length}

Let $T$ be an optimal $r$-ary tree for given $P$ and  $\Lambda = \set{ \gamma_1, \ldots, \gamma_g}$.

All leaves in $T$ are at a  level $\gamma_i$ for some $\gamma_i \in \Lambda$.
Consider any  internal  node $v$ at level $\gamma_{i-1}$. It has no leaf descendants at any level $\ell$ with $\gamma_{i-1} < \ell < \gamma_{i}.$ We may therefore assume that all of its $r^{\gamma_{i}-\gamma_{i-1}}$ descendants at level $\gamma_{i}$
are in the tree, i.e., that $v$ is the root of a {\em complete} subtree of height $\gamma_{i}-\gamma_{i-1}$.

We may therefore create a new tree $T'$ as follows.  The root of $T'$ corresponds to the root of
$T.$  Nodes in $T'$ at level $i$ are in $1$-$1$ correspondence with nodes at level $\gamma_i$ in $T$ and there is
an edge from node $u$ on level $i-1$ to node $v$ on level $i$ in $T'$ if $u$ is the level $\gamma_{i-1}$ ancestor
of $v$ in $T.$  See Figure \ref{Fig:Reserved} for an illustration.

By construction,  $T'$ is a GMR tree with 
$r_i = r^{\gamma_{i} - \gamma_{i-1}}$ and $c_i = \gamma_{i} - \gamma_{i-1}.$  Furtheremore,  the construction
can be reversed,  with any generalized mixed arity tree with these parameters being transformable
into a restricted length tree with the same cost for the given $\Lambda$. 

Since there are at most $g$ levels, our generic GMR algorithm solves this problem
in $O(g n^2)$ time,  improving upon the $O(g n^3)$ algorithm of \cite{baer-2007} .

\vspace*{-.1in}
\subsubsection{The $g$-lengths case}
\label{sec:given-num-length}
If the levels on which leaves appeared were known to be
$\Lambda = \set{ \gamma_1, \ldots, \gamma_g}$
then this is  exactly the 
given-lengths case, which as seen, is equivalent to building an optimal  GMR tree
with $(r_i, c_i) = (r^{\gamma_{i} - \gamma_{i-1}},\gamma_{i} - \gamma_{i-1}).$
The added complication here is to {\em guess} $\Lambda.$

This is equivalent to the problem of building a slightly generalized version of a GMR tree
in which, instead of guessing $\Lambda$, we instead,
at each level $i$, guess  the pair   $(r_i, c_i) = (r^t, t)$ for $t \ge 1.$  
Any such pair is allowable but once $t$ is chosen,  it applies to all nodes
on level $i.$  Furthermore, since the tree only needs $n$ leaves we may assume
$r^t \le r n$ and thus may  restrict $t \le 1+ \log_r n.$

This motivates slightly modifying the GMR model to allow {\em choices}  of $(r_i,c_i)$.

Recall that  the original definition of GMR specifies arities
$R=(r_1, r_2, \ldots)$ and edge lengths $C=(c_1, c_2, \ldots)$.
We now replace  these with 
$\overline R=(\overline r_1, \overline r_2, \ldots)$ and edge lengths $\overline C=(\overline c_1, \overline c_2, \ldots)$
where 
$$
\overline r_i = \{r_{i,1},\, r_{i,2},\, \ldots, r_{i,\Delta_i}\},
\quad
\overline c_i = \{c_{i,1},\, c_{i,2},\, \ldots, c_{i,\Delta_i}\},
$$
are {\em sets}  of $\Delta_i$ possibilities for level $i.$
A {\em permissible tree} is a GMR tree for some sequence
$R=(r_{1,j_1}, r_{2j_2}, \ldots)$ and edge lengths $C=(c_{1,j_1}, c_{2,j_2}, \ldots)$
where $\forall i,\, 1 \le j_i \le \Delta_i$.

Given $P,$ an optimal tree would now be a min-cost permissible tree for
the given $\overline R$, $\overline C$.

The discussion above tells us that to solve the 
$g$-lengths problem, it is only necessary is solve this new generalized version of
the GMR problem to construct  a minimum-cost $g$-level tree
where, for every $i,$  $\overline r_i$ and $\overline c_i$ are the  sets
defined by 
$$\Delta_i = 2+\lfloor \log_r n\rfloor,\quad
\mbox{ and } \quad 
\forall 1 \le j \le \Delta_i, \quad
r_{i,j} = r^{j-1},\ 
c_{i,j} = j-1.
$$ 

The modifications to the definitions and algorithms are straightforward.
Signatures are defined the same way as before. 
Definition \ref{def:valid_1}  of valid signatures needs to be modified to allow
 $r_{i,j}.$ 
 \begin{definition}
 \label{def:valid_2}
$(m,b)$ is a {\em valid $i$-level signature} if
\begin{itemize}
\item If $b >0$ then $m+b \le n.$
\item If $b=0$ then $\exists j,$ such that $\max(n,r_{i,j}) \le m \le n +r_{i,j}-1$
\end{itemize}
\end{definition}
 Note that the number of valid $i$-level signatures is $O(n^2 + n \Delta_{i}).$\\ 
 Definition \ref{def:sig_exp} also needs to be slightly generalized:
 \begin{definition}
\label{def:sig_exp_2}
If $(m',b'),$  $(m,b)$ are, respectively,  valid $(i-1)$ and $i$-level signatures such that  $ 0 \le  b \le b' r_{i,j}$
and $m = m' + b' r_{i,j} - b$, 
we write
$$(m',b) \ijarrow (m,b).$$
\end{definition}

We now, similarly as before, define
  \begin{equation}
  OPT^i[m,b] =  min\left\{Cost_i(T)\,\mid\, \exists T, \textit{T is a i-level tree
      with}\ sig_i(T)=(m,b)\right\}
  \end{equation} 
  
The only major difference is in the analogue of Lemma \ref{lem:DP_basic}, which
gives the DP for calculating $OPT^i[m,b]$.  This now needs to
be split into two phases; the first  calculates, for every $j$,
the optimimum value
of $OPT^i[m,b]$ assuming that $(r_{i},c_{i}) = (r_{i,j}, c_{i,j}).$  The
second takes the minimum of this value over all $j.$  More specifically:
\begin{lemma}
\label{lem:DP_basic2}
\begin{equation}
\mbox{For $1 \le j \le \Delta_{i-1}$, set}
\qquad 
OPT^{i,j}[m,b] = 
\min_{ {\tiny \{(m',b') \,\mid\,  (m',b') \ijarrow (m,b)\}} }
		\Bigl\{ OPT^{i-1}[m',b']+ c_{i,j} W_{m'}\bigr\}.
\end{equation}

The optimal cost of an $i$-level tree with signature $(m,b)$ then 
satisfies
\begin{equation}
OPT^i[m,b] = \min_{1 \le j \le \Delta_i} OPT^{i,j}[m,b]
\end{equation}
Initial conditions are that $OPT^{0}(0,1) =  0$ with all
other entries being set to $\infty$.		
\end{lemma}

Given the values
$OPT^{i-1}[m,b]$, the  batching speedup of Subsection \ref{subsec:batching} now permits,
for any fixed $j$,  calculating all of the values 
$OPT^{i,j}[m,b]$ in $O(n^2)$ time.  Thus,  all of the values  
$OPT^{i}[m,b]$ can be calculated in $O(\Delta_i n^2)$ time.

The total amount of work required for calculating $OPT^{g}[m,b]$ from scratch is then
$O\left(n^2 \sum_{i=0}^g \Delta_i\right)$.  

In the $g$-lengths problem,  $\Delta_i = O(\log_r n)$ so the total runnnig time for
solving the $g$-lengths problem is $O(g n^2 \log n)$, improving the  $O(n^2 g^2 \log^{g-1}n)$
running time of the algorithm in  \cite{baer-2007}.

\vspace*{-.1in}
\section{One-Ended Coding}
\label{sec:One_Ended}

\vspace*{-.05in}
We now consider the problem of constructing minimum-cost
binary prefix-free codes having the property that each codeword ends
with a ``$\bf 1$''.  The original algorithms \cite{BY-90,CSP-94} for this problem were exponential.
\cite{ChanG00} presented a top-down DP running in $O(n^3)$ time.
Using the batched speedup technique developed in Subsection \ref{subsec:batching}
we can  develop a modified top-down DP that reduces the running time to $O(n^2)$

As in Section \ref{Sec:topdown},  the algorithm will find a min-cost coding tree.  We must first
modify the code-tree correspondence to reflect the new $1$-ended requirement.  Assume that 
a left edge is labelled with a `$0$' and a right edge with a `$1$'. A node is a
$0$-node ($1$-node) if the edge connecting it to its parent is labelled by a $0$ ($1$).
We will extend this naturally to $0$-leaves and $1$-leaves, and   $0$-internal nodes and
$1$-internal nodes.

To reflect the $1$-ended restriction on the codes,  only $1$ leaves will be labelled
with probabilities from $P=\{p_1,\ldots,p_n\}$. Let $v_i$ be the $1$-leaf in tree $T$ 
labelled with 
$p_i$.  Then we may still write
$Cost(T) = \sum_i d(v_i) p_i$.  As before,  we pad $P$ so that if $i >n$ then $p_i=0.$

These changes require that we naturally modify the definition of full-trees, signatures and expansions.  
Doing this yields  a DP with a size $O(n^2)$ $OP[m,b]$ table. The naive
algorithm for filling in this table would require $O(n^3)$ time but a batching argument very
similar to that in Subsection \ref{subsec:batching} permits  filling in the table in $O(n^2)$ time.
The step by step modifications required to change the GMR algroithm into one for $1$-ended coding
are given in the Appendix.

\newpage

\bibliographystyle{plain}
\bibliography{Golin_SODA_09_Extended}

\newpage

\appendix
\section{The One-Ended Coding Algorithm}

\begin{figure}[t]
\centerline{\epsfig{file=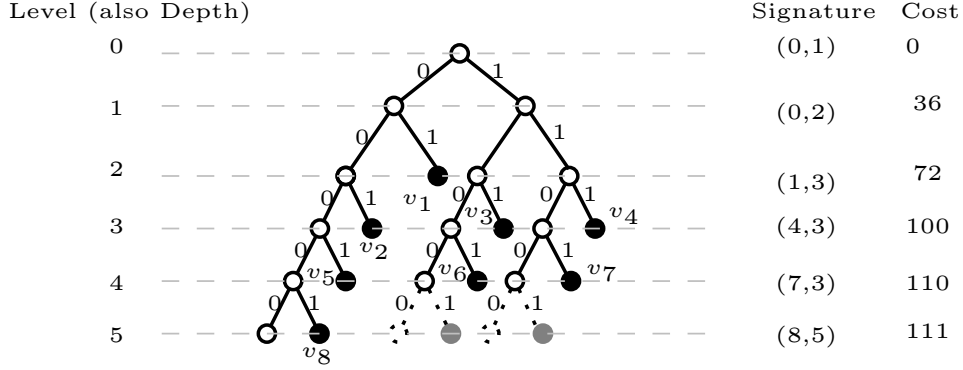, width= 5.0 in}}
\caption{A One-Ended Coding tree for $n=8$ leaves.  Note that only $1$-nodes are labelled leaves. On the bottom level, the two dotted $0$-leaves and their two grey $1$-leaf siblings are `bad' nodes added to make the tree full.
 Signatures and Costs at level $i$ are those of the
(truncated) subtree containing the first $i$ levels of the tree. Costs are calculated for 
 $P=\left\{1,2,3,4,5,6,7, 8\right\}$.}
\label{Fig:One_Ended} 
\end{figure}

We modify the simplifying assumptions on optimal trees $T^*$ 
of Subsection \ref{subsec:Basic TD DP} to reflect
 the extra requirement of being $1$-ended.   Note that not every optimal tree will satisfy
 these assumptions but we will show that at least one optimal tree will.  
 These assumptions will therefore permit us  to restrict the space of trees in which to search.
 Figure \ref{Fig:One_Ended} illustrates the concepts and definitions introduced here.

\medskip

\par\noindent{\em Assumption 1:}  If $i < j$ then, in $T^*$,  $depth(v_i) \le depth(v_j)$.\\
This is the same as before

\medskip
\par\noindent{\em Assumption 2:}  Let $\ell = \ell(T^*).$\\
(a) $T^*$ is full.\\
(b)  
All  $1$-internals are on levels $< \ell-1$.\\
(c) The only $0$-leaves are on level $\ell.$\\
(d) There will be at most $n-1$  $1$-leaves in $T^*$  on levels $<\ell$\\
(e) The number
of $0$-internals on level $\ell-1$ will be   $\le n-1$.

Let $T$ be an optimal $1$-ended tree with exactly $n$  $1$-leaves. Erase all 
$0$-nodes that do not have a $1$-leaf descendant and then  make the tree full
by adding appropriate missing edges so that every internal node has two children.  
Note that after doing this, 
the $1$-sibling of every $0$-leaf exists in $T$ 
 as either an internal node or one of the $n$  $1$-leaves.

Let $v_n$ be the node labelled with $p_n$.  We may
assume that $\ell(v_n)=\ell$;  otherwise we could erase all nodes on level $\ell$
and get a smaller full tree with the same cost which could replace  $T.$

Suppose $v$ is a $1$-internal node that has fewer than two $1$-leaf descendents.  Since
$T$ is full $v$ must have exactly one $1$-leaf descendent.  Erase the subtree rooted at $v$.
 The  resulting tree would still have $n$ $1$-leaves
but a smaller cost,  contradicting the optimality of $T.$ 
Thus every $1$-internal has at least two 1-leaf descendents.  This immediately implies that
all $1$-internals have subtrees of height at least $2$ hanging off of them.
so all $1$-internal nodes are on levels $< \ell-1$.

If $u$ is a $0$-leaf then $\ell(u) \ge \ell -1$;  otherwise,  make $u$ the parent of a $1$-leaf
at level $\ell(u)+1 < \ell$ and move $p_n$ to this $1$-leaf.  This reduces the cost
of the tree,  contradicting the optimality of $T.$  

Thus, all $0$-leaves in $T'$ are on level $\ell-1$ or $\ell$.  
Let $u$ be any $0$-leaf on level $\ell-1$.  Since $T$ is full, its $1$-sibling $v$
is also in $T$,  Since $\ell(v) =\ell-1$, we have already seen that $v$ is a $1$-leaf.

By assumption,  since $T$ has exactly $n$ $1$-leaves and one of them is at level $\ell,$
the number of $1$-leaves on levels $<\ell$ is at most $n-1$.
In particular, since every $0$-leaf on level $\ell-1$ has a $1$-leaf sibling,
there are most $n-1$ such $0$-leaves.

Now create $T^*$ from $T$ by making every $0$-leaf on level $\ell-1$ internal.  This adds 
at most $n-1$ new $1$-leaves to the tree; all of the
new $1$-leaves added will be at level $\ell$ and be labelled with the padded $0$s, 
so this does not change the cost. Since $T$ is optimal for $n$ $1$-leaves,  $T^*$ is optimal for the number of $1$-leaves it has.

The $T^*$ thus created satisfies the assumption.

The remaining two assumptions will be the same as in the GMR case.

\medskip
\par\noindent{\em Assumption 3:} $\level(T^*) \le n$.

\medskip

\par\noindent{\em Assumption 4:}  The optimality of  a tree is fully determined by its {\em leaf sequence},  i.e., the number of
leaves on each level.  No other structural properties need to be maintained. 

\medskip

Our characterization of nodes will be slightly different than in the GMR case.
A node in $T^*$ will be a {\em good } if it's a  $1$-leaf that gets labelled with one of
$p_1,\ldots,p_n.$ 
It will be {\em bad} if it is a $0$-node or a $1$-node that doesn't get lablelled with one of
$p_1,\ldots,p_n$. 
 
We will build trees $T^*$ from the top-down, level-by-level,  At step $i$ our current tree $T$ will
be the first $i$ levels of $T^*.$  We will identify in $T$ the 
 the number of nodes that are good leaves and the number of nodes,  on $T$'s
bottom level,  that are bad.   Note that if $i < \ell(T^*)$ then,  in the next expansion of level,
{\em all} bad nodes on the current bottom level will become internal,  each one contributing one new
$1$-node and one new $0$-node on the new bottom level.
  If $i = \ell(T^*)$ then $T=T^*$ and the bad nodes are 
extra (those that get labelled with $p_j,$ $j >n$) $1$-leaves  and 
$0$-leaves on level $i.$

This motivates us to change the definition of signatures and cost as follows:

\begin{definition}\ \\
\label{def:signature_one_ended}
  If $T$ is  an $i$-level tree its  \emph{$i$-level signature}  is the  ordered
  pair 
  $$sig_i(T)=(m,b)$$ 
  in which 
  \begin{eqnarray*}
  m &=& |\left\{v\in T\,\mid\, \textit{v is a good $1$-leaf, } \level(v) \le i\right\}|\\
  b &=& |\left\{v\in T\,\mid\, \textit{v is a bad node, } \level(v) = i\right\}|
  \end{eqnarray*}
In the above definition, $b$ is counting leaves in $T$ that, if the tree is expanded one level further, 
will become internal nodes.
Let $T$ be the starting ($0$-level) tree containing only the root.  Since the root will always be
expanded, it is bad, so $sig_0(T) = (0,1).$   This will later be the starting point of our dynamic
program.

  Let $T$ be an $i$-level tree with $sig_i(T)=(m,b)$.  The $i$-level
\emph{partial cost} of $T$ is
\begin{equation}
\label{eq:partial-cost_one_ended}
Cost_i(T)=\sum_{t=1}^m depth(v_t)\cdot p_t+ i \cdot\sum_{t=m+1}^n p_t
\end{equation}
  
\end{definition}

\medskip

As in the GMR case, we want to build an optimal  tree $T^*$ that satisfies the assumptions.
We start by noting that, by assumptions 2(b) and 2(c), 
the only bad leaves are on the bottom level $\ell= \ell(T^*).$  

The parents of these bad leaves are on level $\ell-1$ and therefore must be  $0$-nodes, since by assumption 2(b), all 
$1$-nodes on level $\ell-1$ are leaves.  From  assumption 2(e) there are at most $n-1$ such $0$-internals. 
So,  level $\ell$ contains at most $n-1$ bad 1-leaves and $n-1$ bad $0$-leaves.

Now suppose that $T^*$ is being built top-down level-by-level.  Let $T$ be the first $i$ levels
of $T^*$ with $sig_i(T)=(m,b)$.  By definition $m \le n.$ Consider the bad nodes on the bottom level
of $T.$  If $i < \ell$ then every bad node is internal. From the fullness of the tree,  every bad 
node on level $i$ must have a $1$-leaf descendent in $T^*$.  There are at most $2n-1$ ($n$ good and $n-1$ bad)
$1$-leaves in $T$ and each one can appear at most once in some subtree rooted at level $i$ so $b \le 2n-1.$
If $i=\ell$ then $b$ is the number of bad leaves in $T^*$ which is $\le 2(n-1).$  We may therefore
assume that $b \le 2n-1.$

This motivates  an analogue of Definition \ref{def:valid_1}:
\begin{definition}
\label{def:valid_one_ended}
$(m,b)$ is a {\em valid  signature} if
$$m \le n
\qquad\mbox{and}\qquad
b \le 2n-1
$$
\end{definition}
We also modify Definition \ref{def:OPT}:
\begin{definition}
\label{def:OPT_one_ended}
  Let $(m,b)$ be a valid signature. Set
  $OPT[m,b]$ to be the minimum $i$-level partial cost over all $i$ and all $i$-level trees $T$
  with signature $(m,b)$. More precisely
  \begin{equation}
  OPT[m,b] =  min\left\{Cost_i(T)\,\mid\, \exists i,T, \textit{T is a i-level tree
      with}\ sig_i(T)=(m,b)\right\}
  \end{equation}
  If no such tree exists,  set $OPT[m,b] = \infty.$
\end{definition}
Note that if $m=n$, then $OPT[n,b] = \sum_{i=1}^n depth(v_t) \cdot p_t$. Thus,
from, our previous discussion
$$\min \{ OPT^i[n,b] \,\mid\,  1 \le b \le 2n-2\}$$
is the cost of the solution to the one-ended coding problem.

Now suppose that $T'$ is an $(i-1)$-level tree with $sig_{i-1} = (m',b').$  If $T'$ is expanded one
more level then the $b'$ bad nodes on level $i-1$ become internal with each one contributing
one $1$-node and one $0$-node to level $i$.  Between $0$ and $\min(b',n-m')$ of the $1$-nodes can become good
$1$-leaves.  The remaining $1$-nodes and all $b'$ $0$-nodes are bad. The newly created $i$-level tree
has $sig_i(T) = (m,b)$  where
\begin{equation}
\label{eq:one_ended_valid}
m' \le m \le n,
\quad
m = m' + 2b' - b,
\quad
1 \le b' \le b \le 2 b'.
\end{equation}
We therefore define expansions as follows:
\begin{definition}
\label{def:sig_exp_one_ended}
If $(m',b'),$  $(m,b)$ are valid signatures that satisfy  (\ref{eq:one_ended_valid}) 
we write
$$(m',b) \rightarrow (m,b).$$
\end{definition}

The next few steps exactly follow the development of 
Lemmas \ref{lem:sig_conv}, \ref{lem:Sum}, \ref{lem:DP_basic}
and Corollary  \ref{cor:path}
with almost identical proofs,  
so we do not state them explicitly here.  The final
result is that

\begin{equation}
\label{eq:one_ended_DP}
OPT[m,b] = 
\min_{ {\tiny \{(m',b') \,\mid\,  (m',b') \rightarrow (m,b)\}} }
		\Bigl\{ OPT^{i-1}[m',b']+  W_{m'}\bigr\}.
\end{equation}
with initial condition $OPT^0(0,1) =  0$.
Note that $(m',b') \rightarrow (m,b)$ implies either $m' < m$ or $m=m$ and $b = 2 b'$.
So if $(m',b') \rightarrow (m,b)$ then $(m',b')$ is lexicographically smaller than $(m,b)$.
Thus,  the table can be filled in in lexicographic order to correctly calculate the
final values. After filling in the table,  the final solution value will be the minimum
value of $OPT[n,b],$ where $1 \le b \le 2n-2$.  The actual tree corresponding
to this solution can be built by backtracking through the table to find the
sequence of expansions
\begin{equation}
\label{eq:sequence_one_ended}
(m_0,b_0) \rightarrow 
  (m_1,b_2) \rightarrow 
  \cdots	\rightarrow 
  (m_{i-1},b_{i-1}) \rightarrow 
  (n,b)
\end{equation}
that corresponds to a min-cost one-ended tree.

The DP table has $O(n^2)$ entries and each entry requires $O(n)$ time to calculate,
leading to an $O(n^3)$ algorithm for constructing the optimal tree.  An algorithm
with this running time
(based upon a slightly different DP) was given in \cite{ChanG00}.

We now show that it is possible,  using the batching technique earlier introduced
for the GMR, to fill in this table in $O(n^2)$ time.
Define
\begin{definition}
\label{def:one ended old and new sigs}
For $1< d $, define
\begin{eqnarray*}
\cali (d)  &=& \left\{ (m,b)   \,\mid\, m+ b=d,\  0 < b,\ m \le n\right\},\\
\cali'(d) &=& \left\{ (m',b') \,\mid\, m'+2b'=d,\    0 < b,\ m \le n \right\}.
\end{eqnarray*}
\end{definition}
For any fixed $d$  $(m,b) \in \cali(d)$, 
the definition of $\rightarrow$ implies that \\[.01in]
\centerline{
``$(m',b') \rightarrow (m,b)$'' \quad if and only if \quad
``$(m',b') \in \cali'(d)$ with $b/2 \le b' \le b$''.
}\\[.02in]

Note that if $(m,b) \in \cali'(d)$ then $m'+b' < d$ so $(m',b') \in \cali(d')$ with
$d'<d$.  Also,  since $m \le n$ and $b \le 2n-1$, we can bound $d \le 3n-1.$
We may therefore fill in the DP table by filling in all the entries in
$\cali(d)$ as a batch in  the order $d=2,3,4,\ldots 3n-1$.

We now see that for fixed $d$ we can, by batching,  fill in all of the entries in $\cali(d)$ in
$O(d)$ time.  This will allow filling in the entire table in $O(n^2)$ time,  delivering
the promised time reduction.

For fixed $d$ suppose $(m',b') \in \cali'(d)$, i.e., $m' + 2 b' = d$.  Since $ 0 \le m'$ and
$0 < b'$ we have 
$1 \le b' \le d/2.$  For such $b'$
 set  
$$\gamma(b') 
= OPT[m',b'] + W_{m'}
= OPT[d- 2b',b'] + W_{d- 2 b'}.
$$
Note that since $m'+b' < d$, the values $OPT[m',b']$ have already
been calculated and can therefore be looked up.  Therefore,  all of the $\gamma(b')$
can be calculated using a total of $O(d)$ time. 
Then, for $(m,b) \in \cali(d)$, 
\begin{equation}
\label{eq:one_ended_gamma_opt}
OPT[m,b] = \min\left\{ \gamma(b') \,\Big|\, \frac b 2 \le b' \le b \right\}
\end{equation}

Stated this way, this is just a special case of the Range Minimum Query (RMQ) problem.

Given an array $A$ of size $n$,  the RMQ problem is 
to construct a data-structure that,
given two indices $i\le j$, will return the index of a smallest valued item in the subarray
$A[i]\ldots A[j].$
There are known algorithms, e.g., \cite{GabowT83,BenderFPSS05}, for $O(n)$ 
construction of a data-structure that permits $O(1)$ RMQ queries. 
In our case, the array is the $b'$ indexed set of $\gamma(b')$ values.  The array
has size $O(d)$ and its values can be constructed in $O(d)$ time.
\cite{GabowT83,BenderFPSS05} then permit us, in $O(d)$ time,  to construct an
RMQ data structure.  Once this structure is constructed,  each value $OPT[m,b]$ can
be found in $O(1)$ time.  Thus, as promised, all of the $OPT[m,b]$ for $(m,b) \in \cali(d)$
can be batched together and evaluated in  $O(d)$ time

\end{document}